\documentclass[%
 reprint,
 amsmath,amssymb,
 aps,
]{revtex4-1}

\usepackage{graphicx}% Include figure files
\usepackage{dcolumn}% Align table columns on decimal point
\usepackage{bm}% bold math
\usepackage{subfigure}
\begin{document}

\preprint{Arxiv}

\title{Curvature Capillary Migration of Microspheres}% Force line breaks with \\
%\thanks{A footnote to the article title}%
\author{Nima Sharifi-Mood}
%\author{Lu Yao}
%\author{Nima Sharifi-Mood}
%\tnotetext[mytitlenote]{Fully documented templates are available in the elsarticle package on \href{http://www.ctan.org/tex-archive/macros/latex/contrib/elsarticle}{CTAN}.}
%\fntext[myfootnote]{These authors contributed equally to this work}
\author{Iris B. Liu} 
\author{Kathleen J. Stebe}
\email{kstebe@seas.upenn.edu}
%\cortext[mycorrespondingauthor]{Corresponding author}
%\ead{kstebe@seas.upenn.com}
\address{Chemical and Biomolecular Engineering, University of Pennsylvania, Philadelphia, PA, 19104}

%\date{\today}% It is always \today, today,
             %  but any date may be explicitly specified

\begin{abstract}
We address the question: How does capillarity propel microspheres
along curvature gradients? For a particle on a fluid interface, there are two conditions that can apply at the three phase contact line: Either the contact line adopts an equilibrium contact angle, or it can be pinned by kinetic trapping, e.g. at chemical heterogeneities, asperities or other pinning sites on the particle surface. We formulate the curvature capillary energy for both scenarios for particles smaller than the capillary length and far from any pinning boundaries. The scale and range of the distortion made by the particle are set by the particle radius; we use singular perturbation methods to find the distortions and to rigorously evaluate the associated capillary energies. For particles with equilibrium contact angles, contrary to the literature, we find that the capillary energy is negligible, with the first contribution bounded to fourth order in the product of the particle radius and the deviatoric curvature. For pinned contact lines, we find curvature capillary energies that are finite, with a functional form investigated previously by us for disks and microcylinders
on curved interfaces. In experiments, we show microsphere migrate along deterministic trajectories toward regions of maximum deviatoric curvature with curvature capillary energies ranging from $6 \times10^3 - 5 \times 10^4
~k_BT$. These data agree with the curvature capillary energy for the case of pinned contact lines. The underlying physics of this migration is a coupling of the interface deviatoric curvature with the quadrupolar mode of nanometric disturbances in the interface owing to the particle's contact line undulations. This work is an example of the major implications of nanometric roughness and contact line
pinning for colloidal dynamics.
\end{abstract}
\pacs{Valid PACS appear here}% PACS, the Physics and Astronomy
                             % Classification Scheme.
%\keywords{Suggested keywords}%Use showkeys class option if keyword
                              %display desired
\maketitle
\section{Introduction}
Capillary interactions are ubiquitous between particles on fluid interfaces. They trap particles at the interface \cite{Pieranski} and determine their ensuing organization \cite{Marcello, Botto, KralchevskyReview}, allowing particles to be  widely exploited in technological applications such as stabilization of foams and emulsions \cite{Clint}, and in settings as diverse as the food \cite{Food}, pharmaceutical \cite{Pharma}, mineral recovery \cite{Bhargava}, and petroleum industries \cite{Petrol}. For microparticles of radius $a$ at planar interfaces of tension $\gamma$, gravity is irrelevant, as the Bond number $Bo=\frac{\Delta \rho g a^2}{\gamma} \ll 1$, where $g$ is the gravitational acceleration constant, and  $\Delta \rho$ is the density difference between the two fluids on either side of the interface. In this limit, particles with undulated contact lines  distort the interface around them, with an associated excess interfacial area. The deformation fields and interfacial area depend on the relative position of the particles, yielding decreasing capillary energy as particles approach \cite {Stamou}. \\  
\indent In this research we are interested in the behavior of isolated microparticles trapped on curved interfaces. In experiment, microparticles migrate along curvature gradients to sites of high curvature, as has now been observed for microcylinders \cite{Marcello}, microspheres \cite{Blanc}, and microdisks \cite{Disk}.  Theoretically, the curvature capillary energy driving this migration is simply the sum of the surface energies and pressure work for particles at the interface. When particles attach to their host interfaces, they change the interface shape owing to the boundary condition at the contact line where the interface meets the particle.  There are two limits for this boundary condition.  The interface can intersect the particle with an equilibrium contact angle $\theta_0$, determined by the balance of surface energies according to Young's equation \cite{Young}. Alternatively, the contact line can be pinned by kinetic trapping at heterogeneities, roughness or other pinning sites on the particle surface \cite{Stamou,Manoharan,Furst1,Furst2,Carlos,Sepideh}. 
Curvature capillary migration depends on the coupling of the particle-sourced distortion with the host interface shape.  For a particle which is much smaller than the capillary length, far from pinning boundaries, and therefore small compared to the prevailing curvature fields, particle-sourced distortions decay over distances comparable to the particle radius.  This scenario lends itself to analysis by a singular perturbation method\cite{Hinch,Nayfeh}.  In a particle-fixed reference frame defined in the plane tangent to the interface,  the host interface  shape  around the particle can be  locally decomposed into terms proportional to the host interface mean curvature and  deviatoric curvature. By rescaling the governing equations with respect to the particle radius, the shape of the interface around the particle can be found over length scales comparable to this radius.  This approach allows clear treatment of limiting boundary conditions on particle sourced distortions and the ranges of validity of local expansions of the host interfaces.  It also clarifies the appropriate  outer limits on (area or contour) integrals used to evaluate the excess area of the interface created by the particle.  Using this method, we have previously solved the curvature capillary energy for disks with pinned contact lines, which applies equally to spheres with pinned contact lines in the limit of weak undulations\cite{Disk}.  We apply the approach to derive the curvature capillary energy for spheres with equilibrium contact angles. While the curvature capillary energies for this  scenario have been  derived previously and reported to be quadratic in the deviatoric curvature of the interface \cite{wurger,wurger1,Blanc,Fournier}, we find that this term has prefactor zero.  We identify the source of the discrepancy between our result and that published previously.  We predict that spheres migrate by capillarity on constant mean curvature interfaces only if their contact lines are pinned, and that spheres with equilibrium contact lines would have energies several orders of magnitude weaker than is observed in experiment.\\
\indent We perform experiments in which we record the trajectories of polystyrene microspheres at hexadecane-water interfaces with well defined curvature fields. We compare the energy dissipated by particle migration to the curvature capillary energy expressions, and find that the spheres migrate in agreement with the expression for pinned contact lines. \\
\section{Theory}
\begin{figure}
\centering
\includegraphics[width=0.35\textwidth]{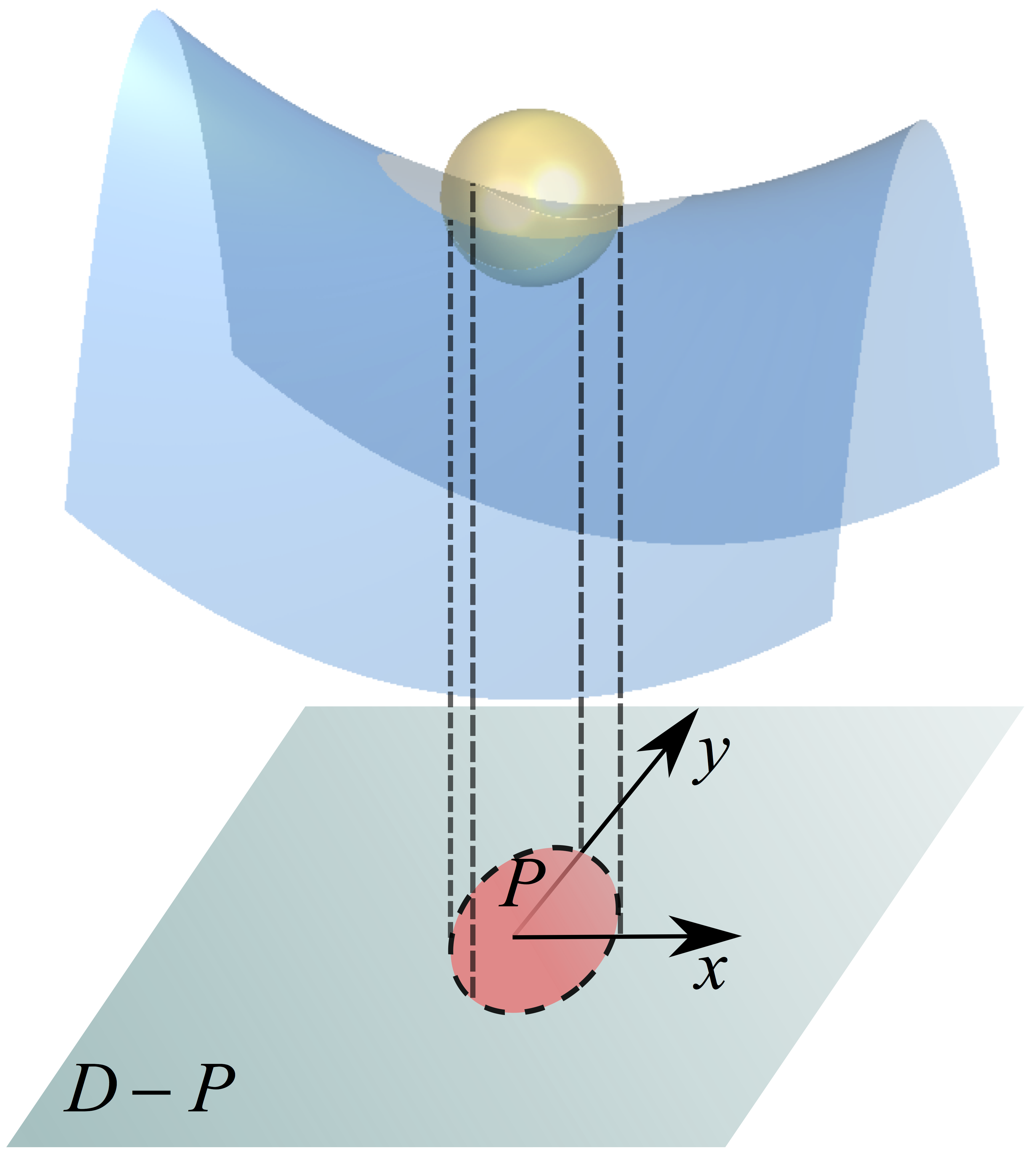}%
\caption{Schematic representation of a sphere trapped at a curved interface and its mapping to the tangent plane.}\label{figs}
\end{figure}
Here we give a concise derivation of the capillary energy of a sphere trapped on a curved fluid interface. Without loss of generality, we focus on interfaces with zero mean curvature as the role of finite mean curvature gradient has been addressed in the literature \cite{wurger1,Disk}.\\
\indent We first consider the free energy of a system including a free interface with a spherical colloidal particle submerged in subphase fluid $1$. In this case, the free energy can be written as, 
\begin{eqnarray}
{E_1} = \gamma \mathop{{\int\!\!\!\!\!\int}\mkern-21mu \bigcirc}\limits_D 
 {(1 + \frac{{\nabla {h_0} \cdot \nabla {h_0}}}{2})dA}+\gamma_1A_s,\label{E_0}
\end{eqnarray}
where $h_0$ is the Monge representation of the host interface height prior to particle attachment, $D$ denotes the entire interfacial domain, $dA=dxdy$, $ \gamma_1$ is the surface energy of the colloid with fluid $1$, and $A_s=4\pi a^2$ accounts for the total area of the colloid. In writing the above expression we have assumed small slopes, i.e. $\left| {\nabla {h_0}} \right| \ll 1$. Upon attaching to the host interface, the particle creates a disturbance to satisfy its boundary condition on the three phase contact line. The free energy of the system in this case is,
\begin{eqnarray}
{E_2} = {\gamma _1}{A_1} + {\gamma _2}{A_2} + \gamma \mathop{{\int\!\!\!\!\!\int}\mkern-21mu \bigcirc}\limits_{D - P} 
 {(1 + \frac{{\nabla h \cdot \nabla h}}{2})dA},
\end{eqnarray} 
where $h$ is the interface height after the particle is adsorbed, ${\gamma _1}{A_1}$ and ${\gamma _2}{A_2}$ are the product of the surface energies and wetted areas for the solid in contact with the upper and lower fluids, respectively, and $P$ denotes the domain under the particle. The capillary energy associated with adsorption of a particle to an interface can be found simply via subtraction of the two energies,
\begin{align}
&E =E_2-E_1={\gamma _1}{A_1} + {\gamma _2}{A_2}-{\gamma _1}{A_s} +  \nonumber \\
&\gamma [\mathop{{\int\!\!\!\!\!\int}\mkern-21mu \bigcirc}\limits_{D - P} 
 {(\frac{{\nabla \eta  \cdot \nabla \eta }}{2} + \nabla \eta  \cdot \nabla {h_0})dA}- \mathop{{\int\!\!\!\!\!\int}\mkern-21mu \bigcirc}\limits_P 
 {(1 + \frac{{\nabla {h_0} \cdot \nabla {h_0}}}{2})dA]},\label{energy}
\end{align}
where $\eta$ is the disturbance field defined as $\eta = h-h_0$. In the above expression, the first integral is due to the disturbance created by the particle and the second integral is the area of the hole in the host interface under the particle. We integrate the first integral by parts and then apply the divergence theorem to have,
\begin{eqnarray}
\mathop{{\int\!\!\!\!\!\int}\mkern-21mu \bigcirc}\limits_{D - P} 
 {[\frac{{\nabla \eta  \cdot \nabla \eta }}{2} + \nabla \eta  \cdot \nabla {h_0}]dA}=\nonumber\\ 
 \oint\limits_{\partial (D-P)} {[\frac{1}{2}\eta \nabla \eta  + \eta \nabla {h_0}] \cdot {\bf{m}}ds},\nonumber\\
\end{eqnarray}
where $\partial (D - P)$ denotes the contours enclosing the domain $D-P$ (See Fig.~1). There are two contours enclosing this domain: one, not shown, infinitely far from the particle, with outward pointing vector in the radial direction, and the other enclosing the region $P$ with outward-pointing unit normal vector $\bf{m}$. Consequently the curvature capillary energy of the system can be expressed as,
\begin{align}
&E =E_2-E_1= {\gamma _1}{A_1} + {\gamma _2}{A_2}-{\gamma _1}{A_s} + \nonumber \\
&\gamma [ \oint\limits_{\partial (D-P)} {[\frac{1}{2}\eta \nabla \eta  + \eta \nabla {h_0}] \cdot {\bf{m}}ds}  
 - \mathop{{\int\!\!\!\!\!\int}\mkern-21mu \bigcirc}\limits_P 
 {(1 + \frac{{\nabla {h_0} \cdot \nabla {h_0}}}{2})dA]}.\label{energy_1}
\end{align}
\indent We expand the interface locally around an arbitrary point to obtain a saddle shape,
\begin{eqnarray}
{h_0}(x,y) = \frac{\Delta c_0}{4}({x^2} - {y^2}),\label{host_1}
\end{eqnarray}
where $(x,y)$ coordinate is tangent to the host interface, oriented along the principle curvatures $c_1$ and $c_2$, and $\Delta c_0$ is the deviatoric curvature of the host interface defined as,
\begin{figure}
\centering
\includegraphics[width=0.45\textwidth]{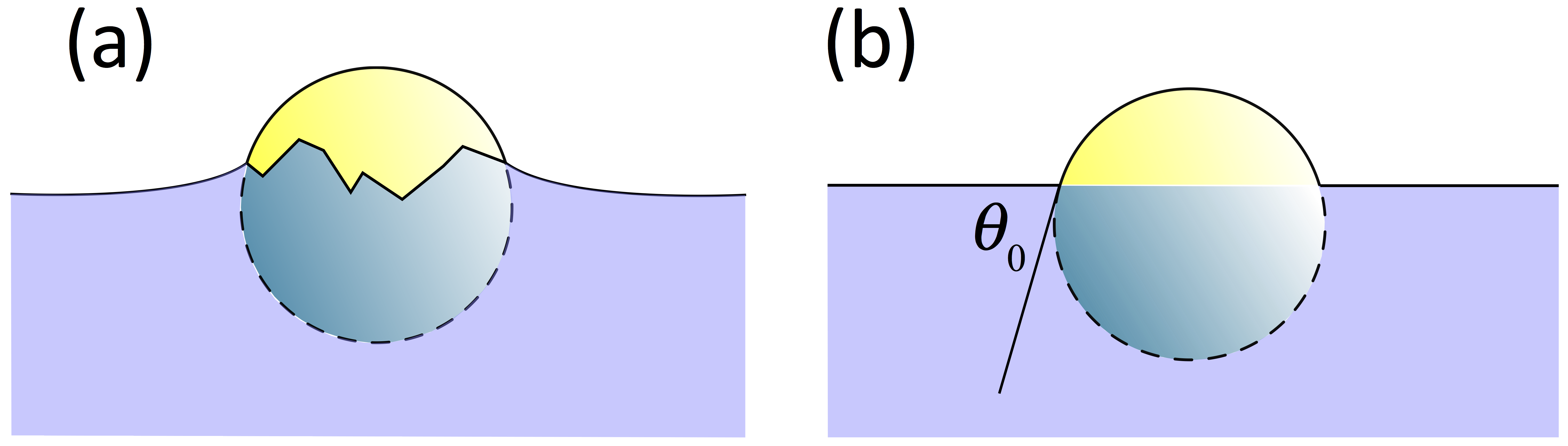}%
\caption{Schematic representation of a sphere trapped at an interface with (a) a pinned undulated contact line and (b) an equilibrium wetting boundary condition with a contact angel $\theta_0$.}\label{fig3}
\end{figure}
\begin{eqnarray}
\Delta c_0=c_1-c_2,
\end{eqnarray}
where we adopt a convention such that $c_1$ is always positive. We discuss the range of validity of Eq.~\ref{host_1} in terms of an asymptotic treatment in Appendix A, and in terms of a comparison of host interface shape used in our experiments  in Fig.~\ref{h_comp}. To evaluate the integrals in Eq.~\ref{energy_1}, we must define the boundary condition at the three phase contact line, and determine the associated disturbance field $\eta$. We discuss two distinct scenarios for this energy (see Fig.~\ref{fig3}).
\subsection{Pinned contact line}
\indent The curvature capillary energy  $E$, for a particle with nearly circular cross section and a pinned contact line trapped on a host interface with arbitrary mean curvature $H_0$ and deviatoric curvature $\Delta c_0$ was  derived previously \cite{Disk}.  The height of the pinned contact line contour can be decomposed into a multipole expansion with quadrupolar mode of amplitude $h_p$. The associated  curvature capillary energy is,
\begin{eqnarray}
E = {E_0} - \gamma \pi {a^2}( \frac{{{h_p}\Delta {c_0}}}{2}+\frac{{3{a^2}H_0^2}}{4} ). \label{energy_capillary}
\end{eqnarray}
In this expression, the first term ${E_0}$ is independent of the local curvature. The second term predicts that a particle will move to sites of high deviatoric curvature, while the third predicts particle migration along gradients of  mean curvature. To understand the relevant importance of these terms, we consider  
$
\frac{{{a^2}H_0^2}}{{{h_p}\Delta {c_0}}} \sim \frac{a}{h_p}\frac{a\Delta p}{\gamma }$ where $\Delta p$ is the pressure jump across the interface (see Appendix~C for non-dimensionalization scheme). This ratio suggests that, for sufficiently small pressure jump across the interface, the effect of mean curvature can be neglected. In section 3, we explore this regime in experiment.  
\subsection{Equilibrium wetting} 
\begin{figure}
\centering
\includegraphics[width=0.45\textwidth]{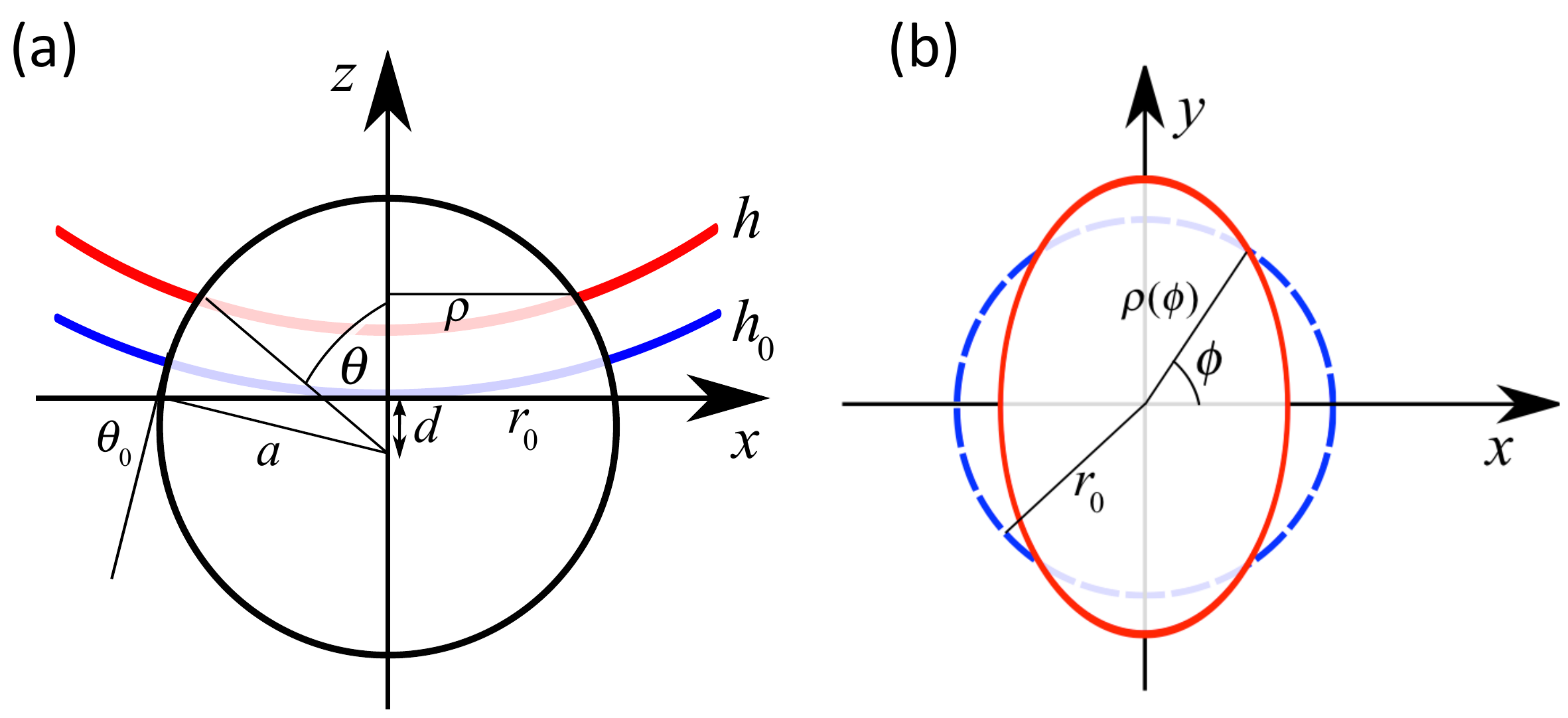}%
\caption{(a) Schematic view of a sphere of radius $a$ at a liquid interface in $x-z$ plane, where $\theta_0$ and $r_0$ are the contact angle and the radius of the contact line for a flat interface. $h_0$ describes a curved interface without a particle, $h = h_0 + \eta$ accounts for the height of interface in presence of the particle. $\theta$ is the polar angle of the deformed contact line, and $\rho$ is its radial position. (b) Schematic view of the contact lines in $x-y$ plane in flat (in blue) and curved interfaces (in red).}\label{fig1}
\end{figure}

\begin{figure}
\centering
\includegraphics[width=0.45\textwidth]{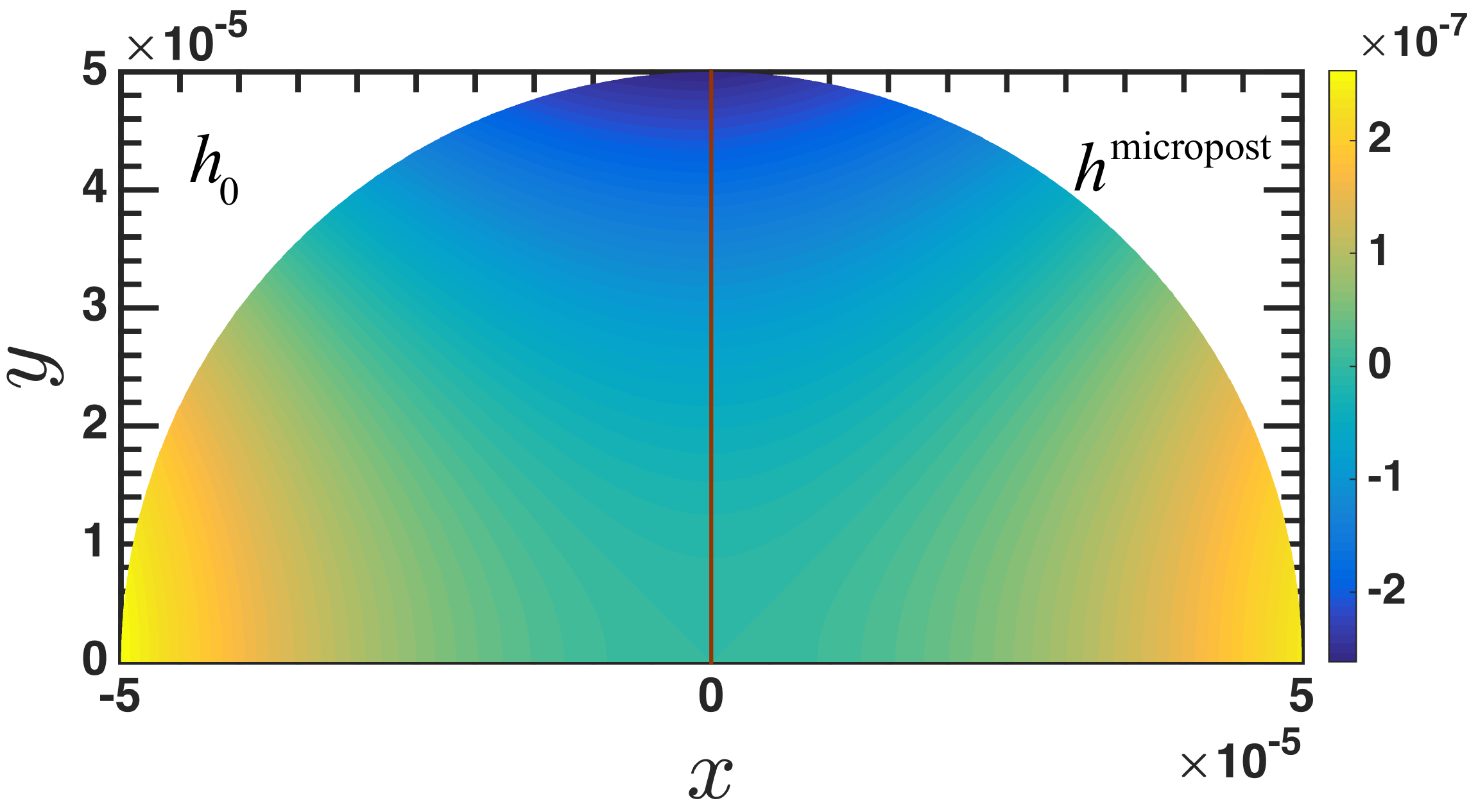}%
\caption{Comparison of the shape of the host interface used in experiment, graphed with respect to the plane at an arbitrary point $L=400~\mu m$ from the micropost center ${h^{\rm{micropost}}}$ (Eq.~\ref{hmicropost}) (right) and Eq.~\ref{far field host}, the limiting form of the interface far from the particle $\mathop {\lim }\limits_{r \to \infty } h(r,\phi )$  (left). The quadrupolar mode of  ${h^{\rm{micropost}}}$ agreees with Eq.~\ref{far field host} to within 2\% error.}\label{h_comp}
\end{figure}

 In Appendix~A, we use a formal matched asymptotic expansion \cite{Hinch,Nayfeh} to evaluate the curvature capillary energy in terms of the small parameter $\lambda=r_0\Delta c_0$, where $r_0=a~\sin \theta_0$ is the radius of the hole made a sphere is a planar interface. Here, we give the main features of this derivation in  dimensional form, and note the bounds on the solutions and ranges of validity implied by the asymptotics.\\ 
\indent For a spherical particle in a curved interface with an equilibrium wetting condition, the fluid interface deforms until it satisfies the equilibrium contact angle at every point on the contact line. The contact line shape is not known {\em {a priori}} and must be determined as a part of analysis, as was originally done  by W\"urger \cite{wurger}. The shape of contact line can be deduced from geometrical relationships  (see Fig.~\ref{fig1}) to be, 
\begin{align}
&\cos \theta  = \cos {\theta _0} + \frac{{h({\bf{r}} \in \rho )}}{{{a}}},\label{1a}\\
&\sin \theta  = \frac{\rho }{{{a}}},\label{2a}
\end{align}
where $\theta$ is a polar angle which located the contact line in a spherical coordinate system with respect to the particle center, $\cos\theta_0=\frac{d}{a}$, and $\rho(\phi)$ is the contour of the contact line projected into a plane, which obeys,
\begin{eqnarray}
\rho  = {a}\sqrt {1 - {{\cos }^2}\theta }.
\end {eqnarray}
\indent By substituting Eq.~\ref{1a} and \ref{2a} in this expression and assuming $\frac{h({\bf{r}} \in \rho )}{a}\sim O(\lambda) \ll 1$, the radial location of contact line projected into the $x-y$ plane  is,
\begin{eqnarray}
\rho  = {r_0} - h({\bf{r}})\cot {\theta _0} - \frac{1}{2}\frac{{{h^2}({\bf{r}})}}{{{r_0}}}({\cot ^2}{\theta _0} + 1) + ...{\left. {} \right|_{{\bf{r}} \in \rho }},\nonumber \\
\end{eqnarray}
\indent The corresponding unit vector $\bf{m}$ and the arc length element $ds$ up to the first correction in $\lambda$ are,
\begin{align}
&{\bf{m}} =  - [{{\bf{e}}_r} - \frac{{\cot {\theta _0}}}{{{r_0}}}\frac{{\partial h({\bf{r}})}}{{\partial \phi }}{{\bf{e}}_\phi }]{\left. {} \right|_{{\bf{r}} \in \rho }},\\
&ds = {r_0}[1 - \cot {\theta _0}\frac{{h({\bf{r}})}}{{{r_0}}}]d\phi {\left. {} \right|_{{\bf{r}} \in \rho }}.
\end{align}
\indent Young's equation requires,
\begin{eqnarray}
{\left. {{{\bf{n}}_P} \cdot {{\bf{n}}_I} = \cos {\theta _0}} \right|_{{\bf{r}} \in \rho }},\label{dot}
\end{eqnarray}
where ${\bf{n}}_P$ is the unit normal to the particle and ${\bf{n}}_I$ is the unit normal to the interface evaluated at the contact line. By applying this expression (see Appendix~A for details), the boundary condition for the interface shape at the three phase contact line to the leading order is,
\begin{eqnarray}
\frac{h}{{{r_0}}} = \frac{{\partial h}}{{\partial r}}{\left. {} \right|_{r = {r_0}}}.\label{three-bc}
\end{eqnarray}
\indent The height of interface satisfies the Young-Laplace equation, which, assuming small slopes, reduces to the Laplacian:
\begin{eqnarray}
{\nabla ^2}h = 0.
\end{eqnarray}
\indent The  boundary condition for the particle-sourced distortion in the far field can be derived using a Van Dyke matching scheme \cite{Hinch,Nayfeh} to match the interface shape far from the particle to the host interface shape. The resulting boundary condition far  from the particle is, 
\begin{eqnarray}
\mathop {\lim }\limits_{r \to \infty } h(r,\phi) = {h_0}=\frac{{\Delta c_0 }}{4}~r^2\cos 2\phi,\label{far field host}
\end{eqnarray}
where the limit $r \to \infty $  implies exploring regions around the particle large compared to the particle radius. Using Eq.~\ref{three-bc} and Eq.~\ref{far field host} the leading order interface shape is,
\begin{eqnarray}
h(r,\phi) = \frac{{\Delta {c_0}\cos 2\phi }}{4}({r^2} + \frac{{r_0^4}}{{3{r^2}}}),
\end{eqnarray}
The corresponding disturbance to the interface is,
\begin{eqnarray}
\eta  = h - {h_0} = \frac{{\Delta c_0\cos 2\phi }}{{12}}\frac{{{r_0^4}}}{{{r^2}}}.\label{eta}
\end{eqnarray}
Asymptotic analysis for the curved interface around the circular microcylinder (Appendix~A) show that Eq.~\ref{far field host} remains valid everywhere the disturbance (Eq.~\ref{eta}) is finite. With this information, we determine the curvature capillary energy Eq.~\ref{energy_1}. In so doing, we can straight-forwardly apply the limits at the contact line and  $r \to \infty $, knowing that these functions remain valid within this domain. This is indeed consistent with experiment. We mold the fluid interface by pinning it to a micropost.  The interface forms a shape with a well defined curvature field. For micro-particles on interfaces with deviatoric curvatures like those in our experiments $\Delta c_0\sim400-3000~m^{-1}$,  the distortions will be $\ll 10^{-10} m$ within $5-6$ radii, $a$, from the the center of particle. Over such ranges, as shown in Fig.~\ref{h_comp}, the shape of the fluid interface Eq.~\ref{far field host} agrees well with the full function describing the interface shape ${h^{\rm{micropost}}}$; the quadrupolar mode of ${h^{\rm{micropost}}}$ agrees with  Eq.~\ref{far field host} to better than $2\%$ error. This is the sole mode that can couple to the quadrupolar undulation of the particle. While we perform this matching in detail only for the host interface used in our experiments, the asymptotic analysis can be applied similarly to any fluid interface provided  $\frac{a}{R_c}\ll 1$, $R_c$ is the characteristic length of the host interface. The deviatoric curvature part of far field boundary condition is given by Eq.~\ref{far field host}, with the details for the curvature source (e.g. the micropost in our example) emerging in the expression for $\Delta c_0$. In the event that the  interface has finite mean curvature, an additional term emerges with the local mean curvature of the interface.\\
  We evaluate\\ 
(i) the self-energy of the disturbance created in the host interface, the sum of the following two contour integrals:
\begin{align}
&\oint\limits_\rho  {\frac{\eta }{2}\nabla \eta  \cdot {\bf{m}}ds = } r_0^4\int_0^{2\pi } {\frac{{\Delta {c_0^2}{{\cos }^2}2\phi }}{{144}}d\phi  = } \frac{{\pi \Delta {c_0^2}r_0^4}}{{144}}\\
&\oint\limits_{r \to \infty } {\frac{\eta }{2}\nabla \eta  \cdot {\bf{m}}ds = } 0\end{align}
This term captures the increase in interfacial area owing to the particle sourced distortion.\\
(ii) the interaction of the disturbance and the host interface, given by the sum of the following two contour integrals:
\begin{align}
&\oint\limits_\rho  {\eta \nabla h_0 \cdot {\bf{m}}ds =  - } r_0^4\int_0^{2\pi } {\frac{{\Delta {c_0^2}{{\cos }^2}2\phi }}{{24}}d\phi  = }  - \frac{{\pi \Delta {c_0^2}r_0^4}}{{24}}\label{4cc}\\
&\oint\limits_{r \to \infty } {\eta \nabla h_0 \cdot {\bf{m}}ds = } r_0^4\int_0^{2\pi } {\frac{{\Delta {c_0^2}{{\cos }^2}2\phi }}{{24}}d\phi  = }  + \frac{{\pi \Delta {c_0^2}r_0^4}}{{24}}.\label{4c}
\end{align}
\indent These terms are equal and opposite, and hence sum to zero. Thus, the net contribution of the particle induced disturbance, the sum of (i) and (ii), is
\begin{eqnarray}
\oint\limits_{\partial (D - P)} {[\frac{1}{2}\eta \nabla \eta  + \eta \nabla {h_0}] \cdot {\bf{m}}ds}  = \frac{{\pi \Delta {c_0^2}r_0^4}}{{144}}.\label{contour_area}
 \end{eqnarray}
(iii) We calculate the energy decrease owing to the area of the hole under the particle as,
 \begin{align}
&-\mathop{{\int\!\!\!\!\!\int}\mkern-21mu \bigcirc}\limits_P 
 {(1 + \frac{{\nabla {h_0} \cdot \nabla {h_0}}}{2})dS = }-\pi r_0^2 -\frac{{\pi \Delta {c_0^2}r_0^4}}{{144}}.\label{hole_area}
\end{align}
\indent The second term in the above expression is equal and opposite to the term in Eq.~\ref{contour_area}. Summing the curvature dependent terms in (i),(ii) and (iii), the energy costs of increased area by the deformation field is shown to be offset by the energy decrease owing to the area eliminated under the particle. This exact canceling of two effects occurs both for the pinned  and equilibrium contact lines. \\
\indent  To complete this derivation, the contribution of the (curvature independent) wetting energies can be simplified by noting, 
\begin{align}
&{\gamma _1}{A_1} = {\gamma _1}2\pi {a^2}(1 + \cos {\theta _0}),\\
&{\gamma _2}{A_2} = {\gamma _2}2\pi {a^2}(1 - \cos {\theta _0}),\\
&{\gamma _1}{A_s} = {\gamma _1}4\pi {a^2},
\end{align}
and 
\begin{eqnarray}
\cos {\theta _0} = \frac{{{\gamma _2} - {\gamma _1}}}{\gamma },
\end{eqnarray}
summing these contributions, the net curvature capillary energy to order $\lambda^2$ is identically zero, i.e.,
\begin{eqnarray}
\frac{{E}}{{\gamma \pi r_0^2}} = E_p + O({\lambda ^4}),\label{energy_wetting}
\end{eqnarray}
where the first term in this expression is Pieranski's trapping energy\cite{Pieranski} for a sphere obeying an equilibrium contact angle in a planar interface 
$E_p =-\sin^{-2}\theta_0(1-\cos\theta_0)^2$. The above expression is exact up to the $\lambda^4$ term; the determination of this bound is described in Appendix~A. \\
  \indent Eq.~\ref{energy_wetting} shows that there is no quadratic term in $\Delta c_0$ for spheres with equilibrium contact lines, and that the curvature capillary energies for these particles are extremely weak. A similar conclusion has been found for perfect disks with circular pinned contact lines\cite{Disk}.  These results differ significantly from prior theory in the literature for this problem. The origin of the discrepancy is an inappropriate treatment of the contour integral given in Eq.~\ref{4c}, which was assumed to be zero in prior work. In Appendix~B this discrepancy is explored in greater detail by evaluating the integrals in questions three ways, all of which agree with our result.  We also follow the approach in the literature for two forms of the contour integral formulation that are, in principle, equivalent, and show that they lead to contradictory results.  
\section{Experiments}
\begin{figure}
\centering
\includegraphics[width=0.45\textwidth]{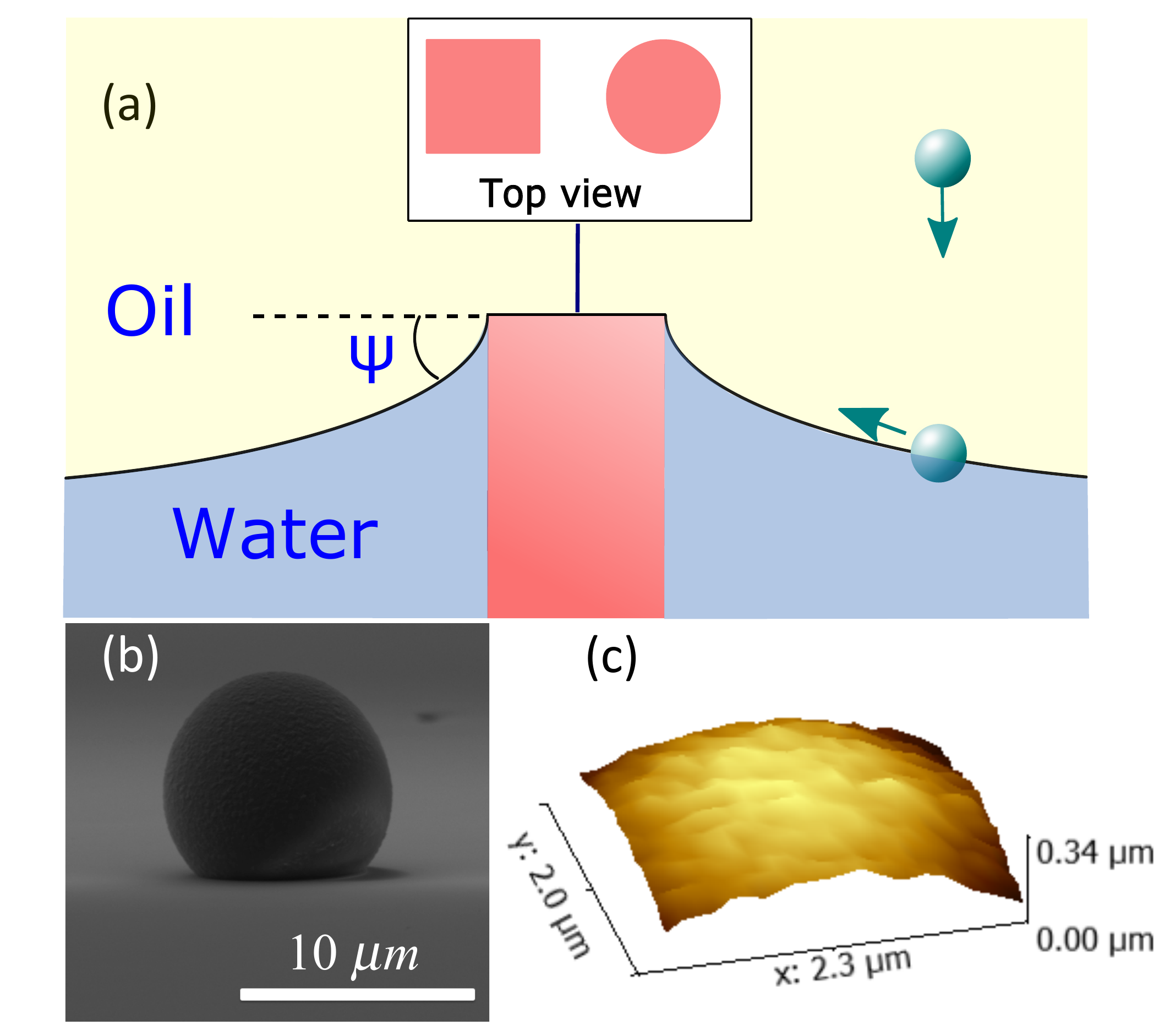}%
\caption{(a) Schematic view of curvature field around the micropost. (b) SEM image of a microsphere and PDMS negative of the air-water interface showing particle roughness and (C) AFM reveals nanoscopic roughness of the sphere surfaces with RMS values ranging of $15 - 21$ nm.}\label{fig2}
\end{figure}

We study migrations of polystyrene colloidal spheres (Polyscicences, Inc.) with mean diameter of $2a = 10~\mu m$. Fig.~\ref{fig2}.(b) illustrates an SEM image of a microsphere revealing qualitatively the surface roughness of the particle. AFM measurement (Bruker Icon) indicates that the root mean squared roughness of the particles is $\sim 15-21~nm$ (see Fig.~\ref{fig2}.(c)).\\
We impose a curvature field to the host interface using a technique reported previously \cite{Marcello} which we recapitulate briefly. A curved oil-water interface is formed around a micropost which is either circular or square in cross section (see the schematic in Fig.~\ref{fig2}(a)). The interface pins to the edge of the post, and has a height $H_m$ at the post's edge. The post is centered within a confining ring located several capillary lengths away $\frac{r_{ring}}{\sqrt{{\frac{\gamma}{\Delta \rho g}}}}=5.5$, where $r_{ring}$ is the radius of the outer ring. By adjusting the volume of water, the slope of the interface at the post's edge is adjusted to be  $\psi \sim 15-18^{\circ}$. This system is gently covered in hexadecane in order to prevent evaporation and to protect the interface from stray convection.  The interface height in a region sufficiently close to the circular post is well approximated by: 
\begin{eqnarray}
{h^{\rm{micropost}}} = {H_m} - {R_m}\tan \psi \ln (\frac{L}{{{R_m}}}),\label{hmicropost}
\end{eqnarray} 
where $L$ is the distance to the center of the post. This interface has zero mean curvature $H_0 = (c_1+c_2)/2 = 0$, and finite deviatoric curvature $\Delta c_0 = c_1 - c_2$ varying with the radial position. Owing to the finite volume of fluid, there is a weak but negligible pressure drop across the interface as we confirm in numerical simulations described in  Appendix~C.\\
\indent A dilute suspension of  microspheres in hexadecane is prepared. A drop of this suspension is carefully dropped on top of the oil phase.  The particles then gently sediment under gravity.  Once attached to the interface, they migrate uphill in a deterministic path along deviatoric curvature gradients. We only focus on isolated particles far from neighbors (distances greater than 10-15 radii) and the micropost ($L>R_m+10a$) to rule out the pair capillary and hydrodynamic interactions \cite{Brenner1961}.
The capillary energy was estimated by evaluating the total dissipation according to the appropriate drag formula (Stokes' law) along particle trajectories. \\
\begin{figure}
\centering
\includegraphics[width=0.45\textwidth]{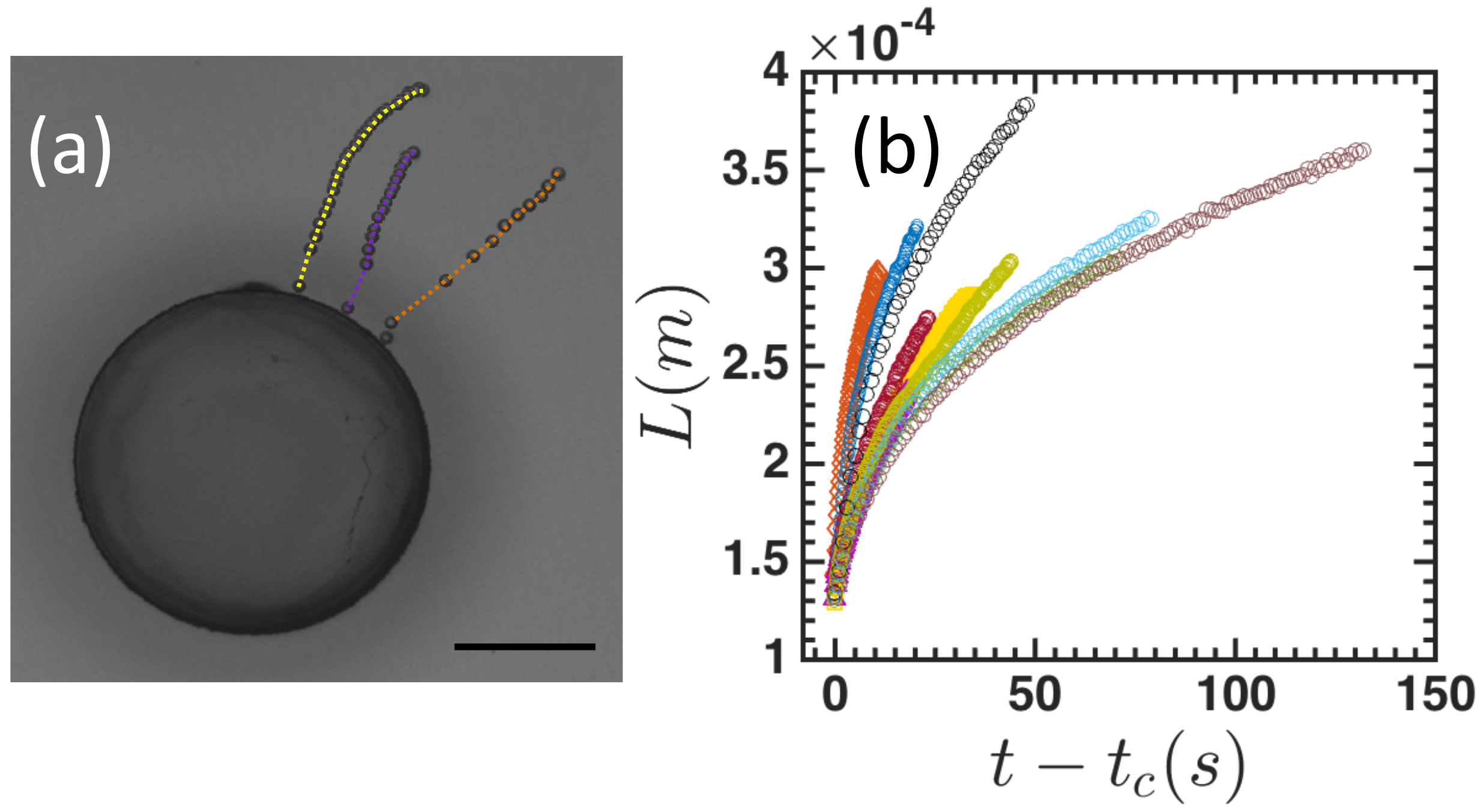}%
\caption{Trajectories of microspheres around a microcylinder. (a) Time stamped image of microspheres trajectory (scale bar = $100~\mu m$). (b) Trajectories for spheres migrating on the interface. Colors for trajectories in (b) correspond to trajectories of similar colors in (a).}\label{fig_traj}
\end{figure}
 \section{Results and Discussions} 
 In Fig.~\ref{fig_traj}(a), we illustrate the time stamped images of trajectories for migrating spheres for constant time increment ($\Delta t = 1~s$). These images reveal that the spheres are propelled faster in the region closer to the post where the  magnitude of deviatoric curvature is greater. Note that the size of the spheres are so small that the inertial effects can be neglected ($Re\sim10^{-3}$) within the entire trajectory. These trajectories are nearly radial, as the corresponding curvature field around the cylindrical post has no dependency on the azimuthal angle $\phi$. Fig.~\ref{fig_traj}(b) shows the radial distance of the migrating microspheres from the center of the post, $L$, as a function of time remaining until contact, $t-t_c$, where $t_c$ is the time in which the sphere reached the edge of the post. Qualitatively, these trajectories are similar to those reported previously for microdisks with pinned contact lines migrating in curvature fields.  To investigate this quantitatively, we compare energy dissipated along the particle trajectory to theory.\\
 \begin{figure}
\centering
\includegraphics[width=0.45\textwidth]{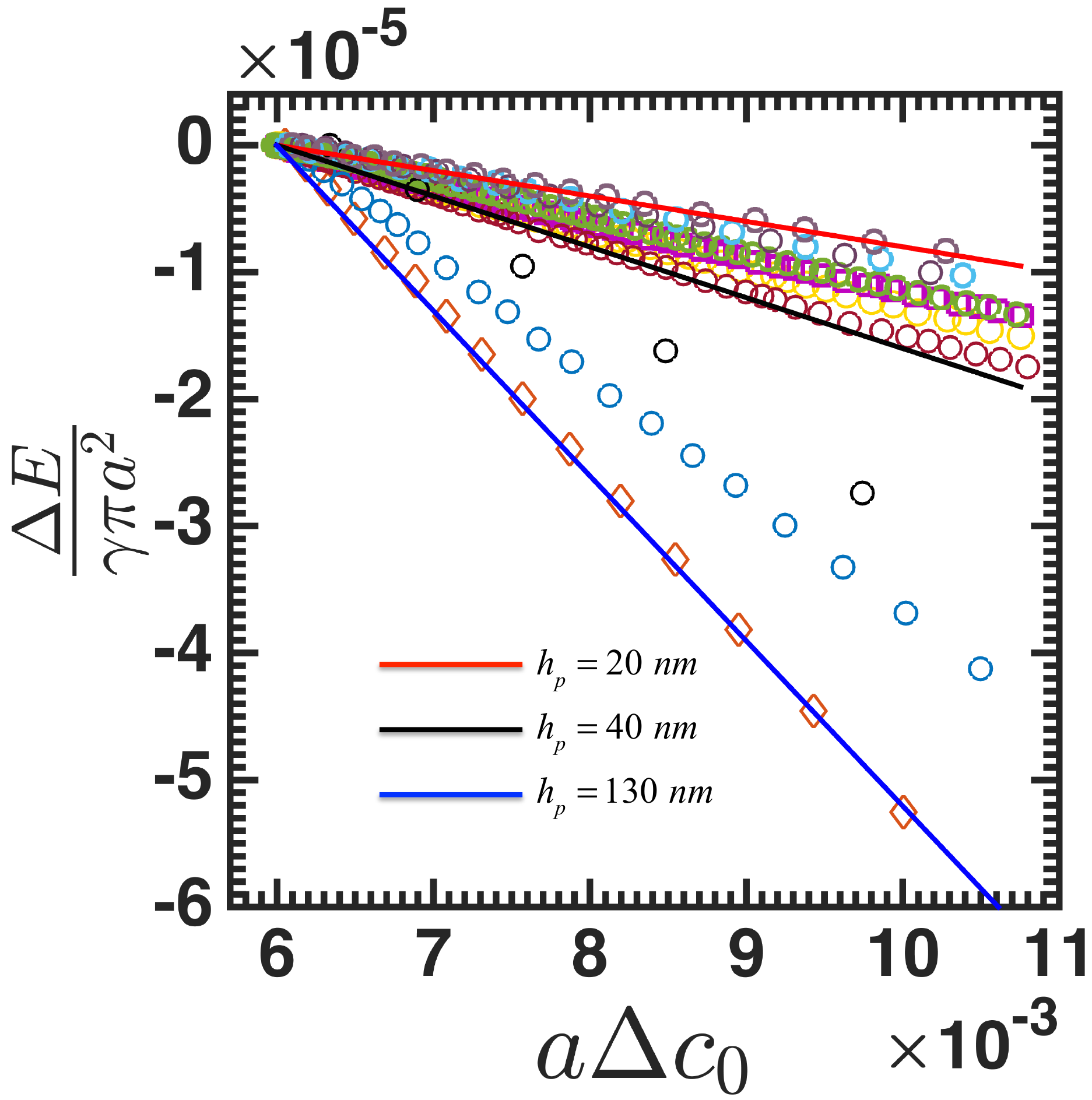}%
\caption{Comparison of predicted curvature capillary energy for particles with pinned contact lines (Eq.~\ref{energy_capillary}) (solid lines) against those extracted from experiment for trajectories in Fig.~\ref{fig_traj}(b) for isolated particles (symbols). The colors correspond to the trajectories in Fig.~\ref{fig_traj}(b). The linear fit is excellent for all trajectories, with coefficient of linear regression ${R^2=0.999}$ for the worst case.}\label{fig_energy}
\end{figure}
     \begin{figure}[h]
\centering
\includegraphics[width=0.45\textwidth]{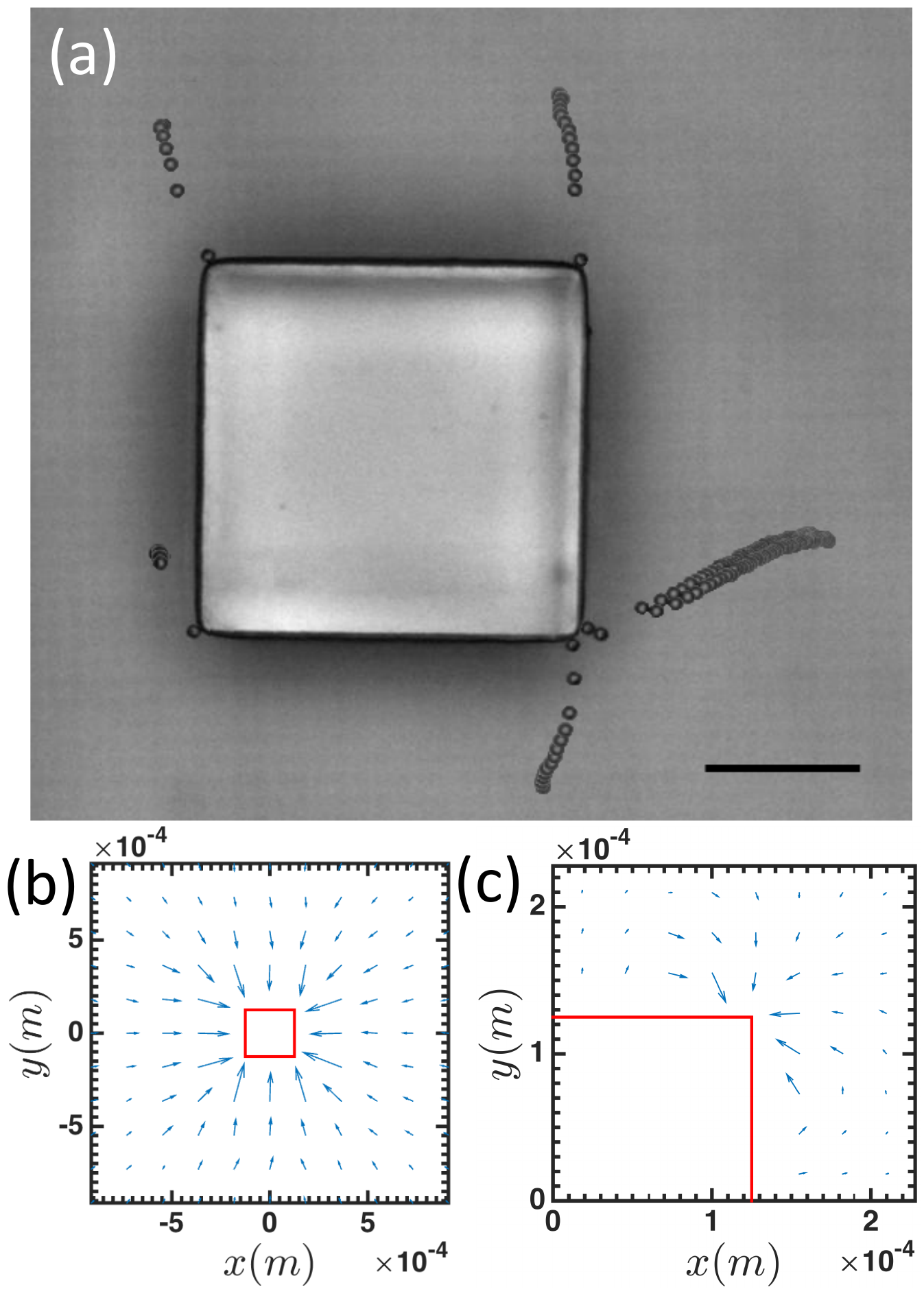}%
\caption{(a) Time-stamped trajectories of microspheres around a square post illustrating that the microspheres follow complex trajectories as defined by the deviatoric curvature field (the scale bar is $100 \mu m$). Numerically evaluated curvature gradient around (b) a square post and (c) a corner of the square post. The arrows scaled with the magnitude of deviatoric curvature gradient. Movies of particle migration are provided as supplementary material.}\label{fig_square}
\end{figure}
 \indent To do so, we note that, in the limit of zero inertia, and neglecting potential energy differences, the curvature capillary energy expended to drive the particles is balanced by viscous dissipation. The total energy dissipated along the trajectories can be extracted from the trajectories according to $\Delta E = \int_{{L_0}}^L {{F_{drag}} dL'}$ where $L_0$ is the reference point and $L$ is an arbitrary point along the trajectory. We used the Stokes' drag formula for a sphere equally immersed in the subphase fluids, $F_{drag}= 6\pi\mu Ua$, where $\mu$ is the average viscosity of oil and water evaluated at the temperature of the environment. The curvature capillary energy found over the trajectories beginning at $a\Delta c_0=6\times 10^{-3}$ and ending ten particle radii from the micropost was plotted against $a\Delta c_0$ as open symbols in Fig.~\ref{fig_energy}; this range of  $a\Delta c_0$ was selected because all trajectories captured in experiment spanned this region. The relationship is linear and the total energy difference along a typical trajectory is thousands of times greater than thermal unit energy $k_BT$. The curvature capillary energy is presented normalized by $\gamma\pi a^2=8.8\times 10^8~k_BT$; the energy for the segment of the particle paths shown in Fig.~\ref{fig_energy} is in the range of $6,000 - 50,000~k_BT$. This magnitude indicates that the equilibrium wetting boundary condition cannot be responsible for the migration of the spheres in our experiment, as according to Eq.~\ref{energy_wetting} for a typical microsphere $a=5 \mu m$ with equilibrium contact angle of $\theta=90^{\circ}$, the curvature capillary energy is of magnitude  $\Delta E \sim \gamma \pi a^2 \sin^2\theta_0 \lambda^4 \sim 10 k_BT$. Moreover the relationship between $\Delta E$ and $r_0\Delta c_0$ would be highly nonlinear.  Hence, we conclude that the equilibrium contact angle boundary condition does not apply to our  microspheres.  Rather, the  microspheres migrate with pinned contact lines.\\
 \indent We propose that this curvature migration is an assay for nanometric corrugations of the contact line in a trapped state. The magnitude of the quadrupolar mode $h_p$ for the contact line undulations can be inferred from the trajectories in Fig.~\ref{fig_traj}. While $7$ of $10$ of the reported trajectories have $h_p$ similar in scale to the particle roughness (between $20-40{\rm{~}}nm$), magnitudes for the remaining trajectories were larger, with  $h_p$ as high as $130{\rm{~}}nm$  for one trajectory.  These results indicate that similar particles have differing pinning states at the interface, with significant consequences for their ensuing dynamics. The role of gravity for particles on curved interfaces has been addressed previously \cite{Blanc,Disk}.  Because the Bond number is negligible, particle weight plays no role in the deformation of the interface, and the analyses above are valid.  For weak enough curvature gradients, however, particles cannot overcome potential energy barriers and thus can attain an equilibrium height.\\
 \indent This form for the curvature capillary energy has been invoked previously to study curvature capillary migration of cylindrical microparticles which followed complex trajectories on interfaces around square microposts with associated complex curvature fields \cite{Marcello}.  If the spheres indeed have identical physics, they, too, should migrate along complex contours in such a curvature field. We have studied trajectories of spheres in this setting (see Fig.~\ref{fig_square}(a), in which isolated particles migrate to corners, as does a pair of dimerized particles at the lower right hand corner). To compare particle trajectories to local curvatures, the curvature field around the square micropost was determined using a Galerkin finite element method, as discussed in  Appendix~C.
   Vectors indicating the magnitude and direction of gradients in deviatoric curvature are indicated in Fig.~\ref{fig_square}(b) and (c). The spheres migrate  from their initial point of attachment along paths defined indeed by these vectors. The ability of the sphere to trace this complex trajectory confirms the underlying physics of microparticles with pinned contact lines is similar regardless of details of the particle shape, and is consistent with the concept that the particle quadrupolar mode couples to the underlying saddle shaped surface.
\section{Conclusion}
We study micropshere migration owing to curvature capillary energy in theory and experiment.   We show that for equilibrium contact angles, the curvature capillary energy is very weak, with leading order contributions of fourth order in deviatoric curvature or higher, in contradiction to the accepted form in the literature. This leading order contribution would amount to roughly $10~k_BT$ in our experiments, several orders of magnitude lower that we measure  in experiment. In  experiment, microspheres migrate along deterministic trajectories defined by curvature gradients.  We  find that the corresponding capillary curvature energy propelling the particles ranges from $6,000-50,000~k_BT$. We compare these observations to arguments derived previously for particles with pinned contact lines, in which the quadrupolar mode of the particle contact line undulation couples with the curvature field to yield an energy linear in the deviatoric curvature. The data indeed obey this form, allowing the magnitude of the particle induced quadrupole to be inferred.  In many cases, it is comparable to particle roughness as determined by AFM.  However, significantly larger magnitudes are also found, suggesting that similar particles can have different pinned states at the interface.  These results imply  that contact line pinning occurs for microparticles at these curved fluid interfaces with dramatic implications in their dynamics at interfaces. 
\section{Appendix~A}\label{apb}
In this section, we provide a detailed derivation for the interface shape when a particle of radius $a$ which attains an equilibrium contact angle is placed on a host interface. The magnitude of this disturbance created by the particle is comparable to the particle radius, $a$. The disturbance is a monotonically decaying function which decays over distances comparable to $a$. Moreover, the particles are small compared to the characteristic length of the host interface $R_c\sim {\raise0.5ex\hbox{$\scriptstyle 1$}\kern-0.1em/\kern-0.15em\lower0.25ex\hbox{$\scriptstyle {\Delta {c_0}}$}}$ and hence the interface domain can be divided into two distinct regions, an inner region close to the particle where the relevant length scale is $a$ and an outer region further away where the relevant length scale is  $R_c$. For small ratio of  $\lambda=\frac{a\sin\theta_0}{R_c}=r_0\Delta c_0\ll 1$, the problem can be solved systematically by a matched asymptotic expansion scheme \cite{Hinch,Nayfeh}. We perform this matching here, in detail, for our particular interface shape:
In the outer region (far from the particle) the shape of interface is known and the shape is given according to a logarithmic expression, 
\begin{eqnarray}
{{\hat h}^{outer}} = \frac{{{H_m}}}{{{R_c}}} - \frac{{{R_m}\tan \psi }}{{{R_c}}}\ln (\frac{{{R_c}\hat L}}{{{R_m}}}),
\end{eqnarray}
please note that the above relation is given in the (outer) coordinate with origin at the center of the circular micropost ($\hat X,\hat Y,\hat Z$) where $\hat Z  =  \hat h^{outer}$ and they are nondimensionalized with respect to the host interface radius of curvature $R_c$.\\
   \begin{figure}
\centering
\includegraphics[width=0.45\textwidth]{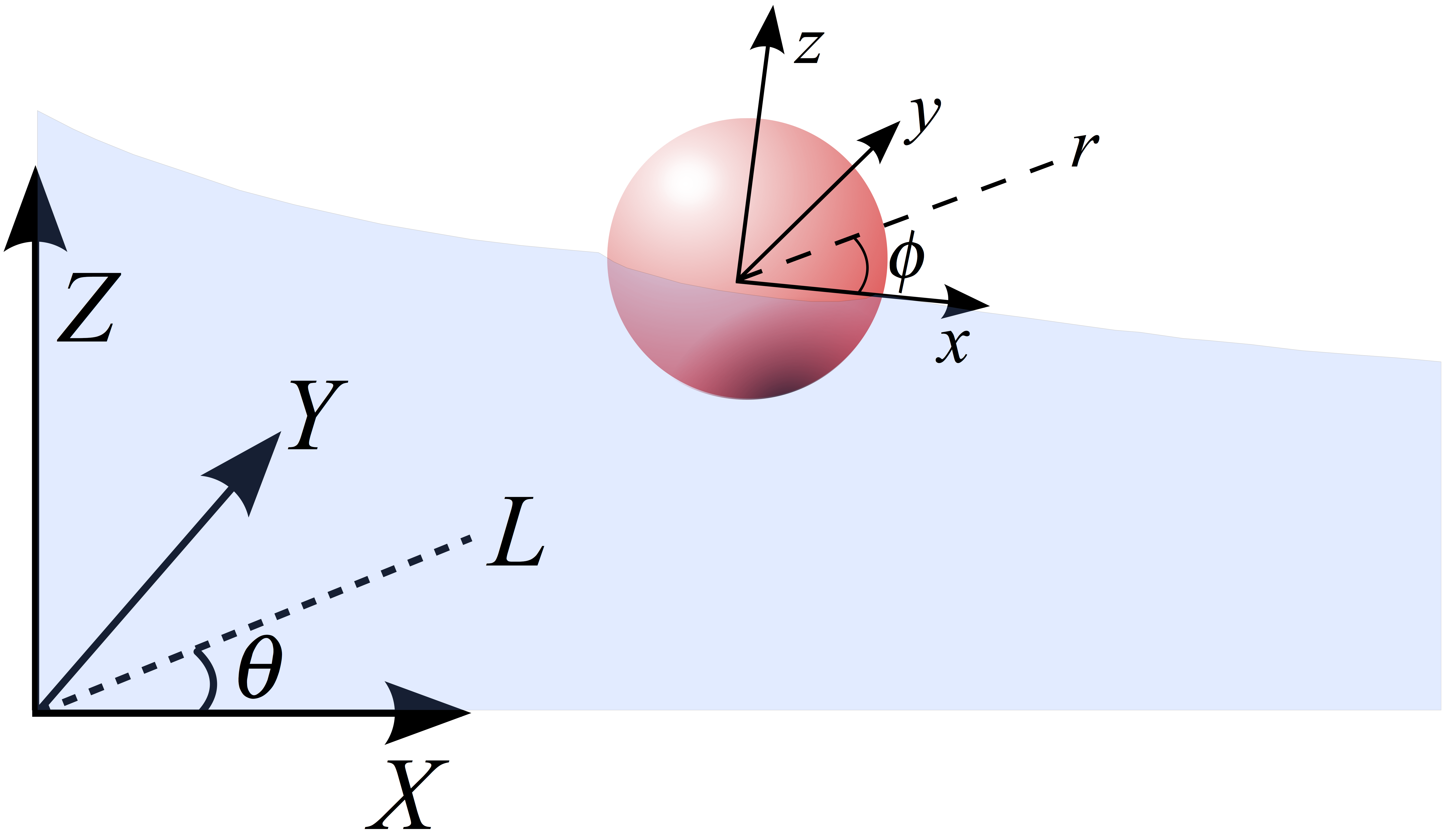}%
\caption{Schematic representation of inner ($x$, $y$, $z$) and outer ($X$, $Y$, $Z$) coordinates.}\label{Fig9}
\end{figure}
In the inner region on the other hand, we adopt inner coordinate $(\tilde{x},\tilde{y},\tilde{z})$ defined with respect to the tangent plane which has slope of $\epsilon=-\frac{R_m\tan\psi}{L_0}$ with respect to the outer coordinate (See Fig.~\ref{Fig9}) and $L_0$ is the location of particle center of mass with respect to the center of micropost.. This coordinates is made dimensionless with inner region length scale $r_0=a\sin\theta_0$. The height of the interface in the presence of the particle is defined $\tilde z = {{\tilde h}^{inner}}$, where we utilize a Monge representation of the interface. The shape of interface in the inner region can be governed via Young-Laplace equation which, in the limit of small gradients and infinitesimal Bond numbers, reduces to:  
\begin{eqnarray}
{{\tilde \nabla }^2}{{\tilde h}^{inner}} = 0.
\end{eqnarray}
We then expand the non-dimensional height of the interface in the inner region as a power series to yield,
\begin{eqnarray}
{{\tilde h}^{inner}} = \tilde{h}_0 + \tilde{h}_1\lambda  + \tilde{h}_2{\lambda ^2} + O({\lambda ^3}).\label{perturb}
\end{eqnarray} 
\indent The equilibrium wetting boundary condition at three phase contact line is dictated by Young's equation,
\begin{eqnarray}
{\left. {{{\bf{n}}_P} \cdot {{\bf{n}}_I} = \cos {\theta _0}} \right|_{{\bf{r}} \in \rho }},\label{dot}
\end{eqnarray}
where ${\bf{n}}_P$ is the unit normal to the particle and ${\bf{n}}_I$ is the unit normal to the interface which can be determined as,
\begin{align}
{{\bf{n}}_P} = {{\bf{e}}_R } = \sin \theta {{\bf{e}}_r} + \cos \theta {{\bf{e}}_z}\label{np}\\
{{\bf{n}}_I} = \frac{{{{\bf{e}}_z} - {\nabla _s}h}}{{\sqrt {1 + {{({\nabla _s}h)}^2}} }},\label{ni}
\end{align}
where ($R,\phi,\theta$) and ($r,\phi,z$) are spherical and cylindrical coordinates located at the particle center of mass, respectively. The operator   $\nabla _s$ is the surface gradient defined as,
\begin{eqnarray}
{\nabla _s} = (I - {{\bf{e}}_z}{{\bf{e}}_z}) \cdot \nabla,
\end{eqnarray}
and $I$ is the unit tensor. By replacing Eq.~\ref{1a} and \ref{2a} in Eq.~\ref{np} and \ref{ni}, Eq.~\ref{dot}  reduces to,
 \begin{align}
&{\left. {{{\bf{n}}_P} \cdot {{\bf{n}}_I}} \right|_{{\bf{r}} \in \rho }} = \frac{{\cos {\theta _0} + \frac{{h({\bf{r}})}}{a} - \frac{\rho }{a}{{\bf{e}}_r} \cdot {\nabla _s}h({\bf{r}})}}{{\sqrt {1 + {{({\nabla _s}h)}^2}} }}{\left. {} \right|_{{\bf{r}} \in \rho }},\nonumber\\
&= \frac{{\cos {\theta _0} + \sin {\theta _0}[\frac{{h({\bf{r}})}}{{{r_0}}} - \frac{\rho }{{{r_0}}}{{\bf{e}}_r} \cdot {\nabla _s}h({\bf{r}})]}}{{\sqrt {1 + {{({\nabla _s}h)}^2}} }}{\left. {} \right|_{{\bf{r}} \in \rho }},\nonumber\\
&= \cos {\theta _0}\{ 1 - \frac{1}{2}[{(\frac{{\partial h({\bf{r}})}}{{\partial r}})^2} + {(\frac{1}{{{r}}}\frac{{\partial h({\bf{r}})}}{{\partial \phi }})^2}]\} {\left. {} \right|_{{\bf{r}} \in \rho }} \nonumber\\
&+ \sin {\theta _0}[\frac{{h({\bf{r}})}}{{{r_0}}} - (1 - \cot {\theta _0}\frac{{h({\bf{r}})}}{{{r_0}}})\frac{{\partial h({\bf{r}})}}{{\partial r}}]{\left. {} \right|_{{\bf{r}} \in \rho }},
\end{align}
where we have utilized the small gradient approximation $\left| {{\nabla _s}h({\bf{r}} \in \rho )} \right| \ll 1$.  Finally, the boundary condition (in dimensional form) at particle contact line can be written as,
\begin{eqnarray}
\frac{{\cos {\theta _0}}}{2}[{(\frac{{\partial h}}{{\partial r}})^2} + {(\frac{1}{r}\frac{{\partial h}}{{\partial \phi }})^2}] = \sin {\theta _0}[\frac{h}{{{r_0}}} - (1 - \cot {\theta _0}\frac{h}{{{r_0}}})\frac{{\partial h}}{{\partial r}}].\nonumber\\
\end{eqnarray}
\indent By recasting this expression in dimensionless form and substituting it in Eq.~\ref{perturb}, we can collect  terms of similar power in $\lambda$.  One can show,
\begin{align}
&{\left. {{{{{\tilde h}_1}}} = \frac{{\partial \tilde h_1}}{{\partial \tilde r}}} \right|_{\tilde r = 1}},\label{bc-contact}\\
&{\left. {{\tilde h}_2 - \frac{{\partial {{\tilde h}_2}}}{{\partial \tilde r}}} \right|_{\tilde r = 1}} =\nonumber \\ &\frac{{\cot {\theta _0}}}{2}{\left. {[{{(\frac{{\partial {{\tilde h}_1}}}{{\partial \tilde r}})}^2} + {{(\frac{1}{{\tilde r}}\frac{{\partial {{\tilde h}_1}}}{{\partial \phi }})}^2} - {{2{{\tilde h}_1}}}\frac{{\partial {{\tilde h}_1}}}{{\partial \tilde r}}]} \right|_{\tilde r =1}},\label{bc-contact-2}
\end{align}
where Eq.~\ref{bc-contact} was derived by W\"urger \cite{wurger}, and is given in Section 2 in dimensional form as Eq.~\ref{three-bc}, the boundary condition at the three phase contact line. Eq.~\ref{bc-contact-2} is the first correction to this contact line boundary condition. In the above, we assumed $\tilde{h}_0=0$.  This 
hypothesis is proven below. The far field boundary condition for the inner solution is obtained by matching the inner solution with the outer solution via Van Dyke's matching procedure\cite{Hinch,Nayfeh}. This can be done systematically by  writing the outer solution in terms of inner variable; thereafter holding the inner variable fixed and expanding for small parameter $\lambda$. This provides the far field boundary condition for the inner region.  Formally, this matching is written:
\begin{eqnarray}
\mathop {\lim }\limits_{\tilde r \to \infty } \lambda {{\tilde h}^{inner}}(\tilde r,\phi ) = \mathop {\lim }\limits_{\scriptstyle\lambda  \to 0\hfill\atop
\tilde {r}~{\rm{fixed}}\hfill} {{\hat h }^{outer}}(\tilde r,\phi ),\label{matching}
\end{eqnarray}
note that in the above expression, $\hat h^{outer}$ is made dimensionless with outer region length scales, and ${{\tilde h}^{inner}}$ is made dimensionless with inner region length scales. \\
 We demonstrate the execution of this matching procedure for our particular host interface, without loss of generality.  The same procedure can be applied to any host interface provided $\frac{a}{R_c} \ll 1$. To perform this matching, we perform a coordinate transformation between the inner and the outer coordinates. The inner coordinate is tangent  with respect to the outer coordinate with slope $\epsilon$. The relationship between the two coordinates providing small slopes $|\epsilon|  \ll 1$ (in dimensional form) can be written as,
\begin{align}
&X = {L_0} + x + O(\epsilon ),\nonumber\\
&Y = y,\\
&Z = {Z_0} + z + O(\epsilon ),\nonumber
\end{align}
where $Z_0$ is the interface height at the particle center of mass with respect to the outer coordinate. Utilizing the above transformation, the right hand side of matching condition expression in Eq.~\ref{matching} can be written as,
\begin{eqnarray}
\hat h^{outer} = \frac{{{H_m}}}{{{R_c}}} - \frac{{{R_m}\tan \psi }}{{{R_c}}}\ln (\frac{{\sqrt {{{(x + {L_0})}^2} + {y^2}} }}{{{R_m}}}) - \frac{{{H_0}}}{{{R_c}}}.\nonumber\\
\end{eqnarray}
By nondimensionalizing and expanding the argument of the $\log$ term for small $\lambda$, one can show,
\begin{eqnarray}
\mathop {\lim }\limits_{\scriptstyle\lambda  \to 0\hfill\atop
\tilde{r}~{\rm{fixed}}\hfill} {{
\hat h }^{outer}}(\tilde r,\phi ) = \frac{{{\lambda ^2}}}{4}{{\tilde r}^2}\cos 2\phi  + O(\epsilon {\rm{,}}{\lambda ^3}),
\end{eqnarray} 
utilizing  Eq.~\ref{matching}, this can be recast to specify the far field boundary condition for the disturbance in the inner region, 
 \begin{eqnarray}
 \mathop {\lim }\limits_{\tilde r \to \infty } {\tilde h^{inner}}(\tilde r,\phi ) = \frac{\lambda }{4}{\tilde r^2}\cos 2\phi+ O(\epsilon {\rm{,}}{\lambda^2}),
 \end{eqnarray}
in dimensional form, this becomes Eq.~\ref{far field host}. This matching also clarifies the meaning of taking the limit  $r\to\infty$. This implies exploring regions around the particle large compared to the particle radius. 
To apply the boundary condition, we introduce a vertical shift factor $\omega_0=\frac{\lambda^2\cot\theta_0}{12}$ to the particle center of mass and therefore we have, 
\begin{align}
&{{\tilde h}^1} = \frac{{\cos 2\phi }}{4}({{\tilde r}^2} + \frac{1}{3\tilde r^2}),\\
&{{\tilde h}^2} =  - \frac{{\cot {\theta _0}~}}{{36}}\frac{{\cos 4\phi }}{{{{\tilde r}^4}}},
\end{align}
This is the inner solution to the shape of the interface up to $\lambda^2$. \\
To construct a uniformly valid (uv) solution over the entire domain, we define a uniformly valid solution in dimensional form according to,
\begin{eqnarray}
{h^{\rm{uv}}} = {R_c}{{
\hat h }^{outer}} + a{{\tilde h}^{inner}} - {R_c}\mathop {\lim }\limits_{\scriptstyle\lambda  \to 0\hfill\atop
\scriptstyle\tilde r~{\rm{fixed}}\hfill} {{\hat h }^{outer}},
\end{eqnarray}
consequently, the (dimensional) disturbance will be,
\begin{align}
&\eta  = {h^{{\rm{uv}}}} - {R_c}{{\hat h }^{outer}} =\lambda\frac{{r_0 }}{{12\tilde r^2}}\cos 2\phi  -\lambda ^2 \frac{{r_0\cot {\theta _0}}}{{36{\tilde r}^4}}{\cos 4\phi},\label{dist_app}
\end{align}
the disturbance is solely a decaying function of inner variable $\tilde {r}$. This implies that its value is identically zero in the outer region. Thus, the particle results in a ``local'' disturbance which fades over a length scale comparable to the inner length scale $r_0$. Furthermore, in evaluating the area owing to the particle disturbance Eq.~\ref{energy_1}, there is no coupling  between $\tilde h^1$ and $\tilde h^2$ owing to orthogonality of $\cos n \phi$.  Therefore, there is no term in the energy expression of order $\lambda^3$.  In the text, we evaluate the energy up to  $\lambda^2$ and find this term to have prefactor zero.  Hence, the leading order contribution to the curvature capillary energy for spherical particles with equilibrium contact lines is order $\lambda^4$.
\section{Appendix~B}
According to the asymptotic discussion, all area integrals for the free energy can be decomposed to two contributions in the inner and outer regions:
\begin{eqnarray}
\mathop{{\int\!\!\!\!\!\int}\mkern-21mu \bigcirc} 
 {fdA}  = \mathop{{\int\!\!\!\!\!\int}\mkern-21mu \bigcirc}\limits_{inner} 
 {{f^{inner}}dA}  + \mathop{{\int\!\!\!\!\!\int}\mkern-21mu \bigcirc}\limits_{outer} 
 {{f^{outer}}dA}.
\end{eqnarray}
In particular the integral which accounts for the interaction of the disturbance with the host interface can be written:
\begin{align}
&\mathop{{\int\!\!\!\!\!\int}\mkern-21mu \bigcirc}\limits_{D - P} 
 {\nabla {h_0} \cdot \nabla \eta dA}  = \nonumber\\
&\mathop{{\int\!\!\!\!\!\int}\mkern-21mu \bigcirc} 
 {{{(\nabla {h_0} \cdot \nabla \eta )}^{inner}}dA}  + \mathop{{\int\!\!\!\!\!\int}\mkern-21mu \bigcirc} 
 {{{(\nabla {h_0} \cdot \nabla \eta )}^{outer}}dA}, \label{int-asypm}
\end{align}
since, as shown in Appendix~A, the disturbance decays to zero in the inner region, the second integral in the right hand side of Eq.~\ref{int-asypm} is zero. Asymptotics also establish that the description of the host interface, $\mathop {\lim }\limits_{r \to \infty } h_0={h_0}^{inner} = (\frac{{{r_0}^2\Delta {c_0}}}{4})\frac{{{r^2}}}{{{r_0}^2}}\cos 2\phi$ remains valid everywhere that the disturbance is finite. This allows the first integral on the right hand side of Eq.~\ref{int-asypm} be evaluated. In the ensuing discussion, the superscript ``inner'' is removed, while understanding that the sole surviving integral is in fact in the inner region. Substituting $h_0$ and $\eta$, the integral can be evaluated directly. Or, the divergence theorem can be used to recast this integral in terms of contour integrals that bound the domain; there are two equivalent forms for this approach.  We evaluate the integral all three ways: \\
(i) Direct integration over the angle $\phi$ shows this integral to be zero at all radial locations between $r_0$  and any location $r^{\ast}$ in the domain.
\begin{align}
&\int_{{r_0}}^{{r^ * }} {\int_0^{2\pi } {[\frac{{\partial {h_0}}}{{\partial r}}\frac{{\partial \eta }}{{\partial r}} + \frac{1}{{{r^2}}}\frac{{\partial {h_0}}}{{\partial \phi }}\frac{{\partial \eta }}{{\partial \phi }}]rd\phi dr} } \nonumber\\
&= {\rm{ }}\int_{{r_0}}^{{r^ * }} {\frac{{\Delta {c_0}^2}}{{12}}\frac{{{r_0}^4}}{r}dr\int_0^{2\pi } {( - {{\cos }^2}2\phi  + {{\sin }^2}2\phi )} d\phi}\nonumber\\
&= \frac{{{r_0}^4\Delta {c_0}^2}}{{12}}\ln (\frac{{{r^*}}}{{{r_0}}})(0) = 0.\label{area_1}
\end{align}  
Thus, for any ÒsliceÓ of the interface around the particle, this integral is zero. One might at first be concerned about the logarithmic divergence in Eq.~\ref{area_1}, but the factor multiplying it is identically zero for any value of $r^{\ast}$.\\
(ii) One form of integration by the Green's theorem. The integrand can be recast as: $\nabla {h_0} \cdot \nabla \eta  = \nabla  \cdot (\eta \nabla {h_0}) - (\eta {\nabla ^2}{h_0})$, where the second term on the right hand side is zero since $\nabla^2 h_0=0$. The area integral can then be recast:
\begin{eqnarray}
\mathop{{\int\!\!\!\!\!\int}\mkern-21mu \bigcirc}\limits_{D - P} 
 {\nabla {h_0} \cdot \nabla \eta dA}  = \mathop{{\int\!\!\!\!\!\int}\mkern-21mu \bigcirc}\limits_{D - P} 
 {\nabla  \cdot (\eta \nabla {h_0})dA},
\end{eqnarray}
the integrals can be written via application of Green's theorem as contour integrals: 
\begin{align}
&\mathop{{\int\!\!\!\!\!\int}\mkern-21mu \bigcirc}\limits_{D - P} 
 {\nabla  \cdot (\eta \nabla {h_0})dA}  \nonumber \\
&= \mathop {\lim }\limits_{r* \to \infty } \oint\limits_{r = r*} {{{\bf{e}}_r} \cdot (\eta \nabla {h_0}}) r~d\phi  - \oint\limits_{r = {r_0}} {{{\bf{e}}_r} \cdot (\eta \nabla {h_0}}) r~d\phi,\label{dd4}
\end{align} 
which ultimately yields the results given in Eq.~\ref{4cc} and \ref{4c}.\\
(iii) Another form of integration by Green's theorem. The integrand can also be recast as $\nabla {h_0} \cdot \nabla \eta  = \nabla  \cdot ({h_0}\nabla \eta ) - ({h_0}{\nabla ^2}\eta )$, where the second term on the right hand sides is zero since ${\nabla ^2}\eta  = 0$.   The area integral is then,
\begin{eqnarray}
\mathop{{\int\!\!\!\!\!\int}\mkern-21mu \bigcirc}\limits_{D - P} 
 {\nabla {h_0} \cdot \nabla \eta dA}  = \mathop{{\int\!\!\!\!\!\int}\mkern-21mu \bigcirc}\limits_{D - P} 
 {\nabla  \cdot (h_0 \nabla {\eta})dA},\label{ee1}
\end{eqnarray}
which, via application of Green's theorem as:
\begin{align}
&\mathop{{\int\!\!\!\!\!\int}\mkern-21mu \bigcirc}\limits_{D - P} 
 {\nabla  \cdot (h_0\nabla {\eta })dA}  \nonumber \\
&= \mathop {\lim }\limits_{r^{\ast} \to \infty } \oint\limits_{r = r^{\ast}} {{{\bf{e}}_r} \cdot (h_0 \nabla {\eta}}) rd\phi  - \oint\limits_{r = {r_0}} {{{\bf{e}}_r} \cdot (h_0 \nabla {\eta}}) rd\phi,
\end{align} 
the two contour integrals are:
\begin{eqnarray}
\mathop {\lim }\limits_{r* \to \infty } \oint\limits_{r = r*} {{{\bf{e}}_r} \cdot (\eta \nabla {h_0})rd\phi }  =  - \frac{{\pi {r_0}^4\Delta {c_0}^2}}{{24}},\\
-\oint\limits_{r = {r_0}} {{{\bf{e}}_r} \cdot {h_0}\nabla \eta } rd\phi  =   \frac{{\pi {r_0}^4\Delta {c_0}^2}}{{24}},\label{dd5}
\end{eqnarray}
The terms are equal and opposite and hence the integral is zero. In the literature, this area integral was evaluated by first recasting it in contour integrals as in Eq.~\ref{dd4}. Thereafter, the far field contour was neglected, and the quadratic term in $\Delta c_0$ was retained as the value for the integral (see Eq.~\ref{4cc}). If the same logic is applied to the area integral as recast in Eq.~\ref{ee1}, only Equation Eq.~\ref{dd5} would be retained. These two forms, which should be equivalent, are in fact equal in magnitude and opposite in sign. This clearly demonstrates that the neglect of the outer contour is improper, as it represents a failure to close the bounding contours on the integral, and violates the Green's theorem used to recast the area integral.     
\section{Appendix~C}\label{ap}
In this section we discuss the numerical method used to calculate the height of interface around the square post. The height of interface in this case can be governed via the linearized Young-Laplace equation according to,
\begin{eqnarray}
\gamma{\nabla ^2}h = \Delta \rho gh + \Delta p,\label{ylp}
\end{eqnarray}
where $ \Delta \rho$ is the density difference of subphase fluids and $\Delta p$ is the pressure jump across the interface.  We scaled the variables as follows,
\begin{eqnarray}
\tilde h = \frac{h}{{\frac{{\left| {\Delta p} \right|}}{{\Delta \rho g}}}},~~~~~~~~~\tilde \nabla  = {(\frac{\gamma }{{\Delta \rho g}})^{\frac{1}{2}}}\nabla.
\end{eqnarray}
Eq.~\ref{ylp} can be rewritten as,
\begin{eqnarray}
{{\tilde \nabla }^2}\tilde h - \tilde h =  \pm 1,\label{ylpn}
\end{eqnarray}
where $\pm 1$ on the right hand side stands for negative and positive pressure jumps across the interface. To obtain the discretized form of Eq.~\ref{ylpn} in the weak form to implement the finite element algorithm for solution, we first multiply both sides of the Eq.~\ref{ylpn} by a weighting function $\varphi_i$ and then integrate over the area of element,
\begin{eqnarray}
\mathop{{\int\!\!\!\!\!\int}\mkern-21mu \bigcirc} 
 {{\varphi _i}{{\tilde \nabla }^2}\tilde hdA}  - \mathop{{\int\!\!\!\!\!\int}\mkern-21mu \bigcirc} 
 {{\varphi _i}\tilde hdA}  =  \pm \mathop{{\int\!\!\!\!\!\int}\mkern-21mu \bigcirc} 
 {{\varphi _i}dA}.
\end{eqnarray}
 Having utilized integration by parts and the divergence theorem, one can show,
 \begin{align}
 &\oint {{\varphi _i}\tilde \nabla \tilde h \cdot {\bf{m}}ds}  - \mathop{{\int\!\!\!\!\!\int}\mkern-21mu \bigcirc} 
 {\tilde \nabla {\varphi _i} \cdot \tilde \nabla \tilde hdA}  - \mathop{{\int\!\!\!\!\!\int}\mkern-21mu \bigcirc} 
 {{\varphi _i}\tilde hdA}  =\nonumber\\  
 &\pm \mathop{{\int\!\!\!\!\!\int}\mkern-21mu \bigcirc} 
 {{\varphi _i}dA},
 \end{align}
   \begin{figure}
\centering
\includegraphics[width=0.45\textwidth]{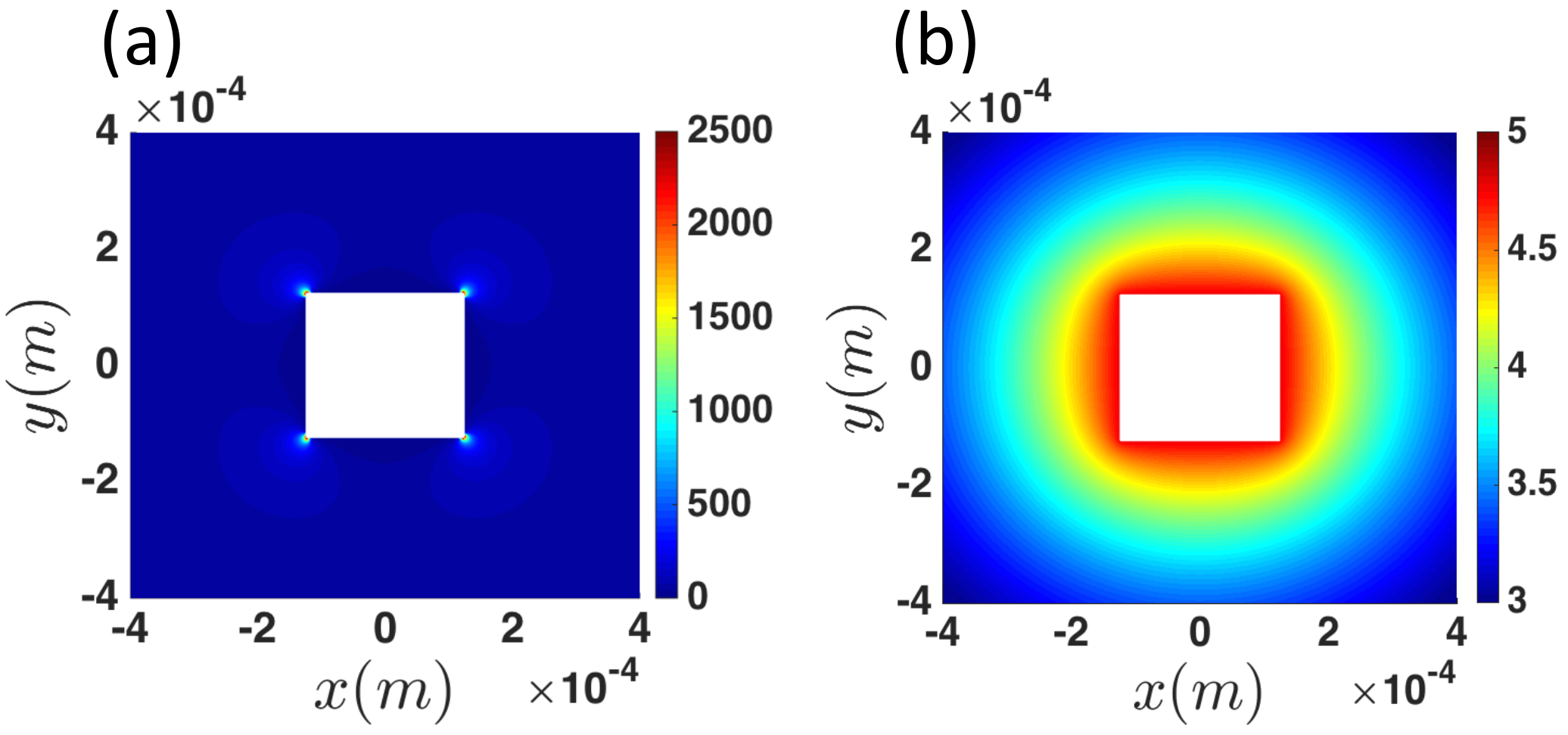}%
\caption{Numerical solution of the curvature field via finite element analysis around a square post. (a) Deviatoric curvature and (b) mean curvature. The magnitudes of the deviatoric and mean curvatures are in $m^{-1}$.}\label{Fig7}
\end{figure}
where $\bf{m}$ is the unit normal vector of the enclosing contours which are pointing outward from the domain of interest. In the above expression, the first term only applies to the boundaries and therefore we neglect it since we have Dirichlet boundary conditions in our problem. Utilizing the Galerkin method \cite{FEM}, the non-dimensionalized height $\tilde h$ were approximated by a piecewise quadratic polynomial function where $\varphi_i=1$ at vertex $i$ and $\varphi_i=0$ at all other vertices. Thus, the non-dimensional height can be approximated as $\tilde h = \sum\limits_{j = 1}^N {{\varphi _j}{{\tilde h}_j}} $, where $N$ is the total number of vertices in the mesh and ${\tilde h}_j$ is the value of the height at the $j$th vertex. Since the problem is 2D, we generated a triangular structured mesh by discretizing the domain.  This was done in such a way that more nodes were available in the transition region (sharp corner) to capture the rapid change in height. The pressure jump was not known {\em{a priori}}; it is found from the slope condition at the middle of the square post.  To do so, we solve the entire problem in an iterative manner. We guess a value for the pressure jump across the interface, solve the problem by applying the Dirichlet boundary conditions at the post and the outer ring, and subsequently calculate the slope at the middle of the square post. The slope is compared to the measured value in the experiment with $\psi = 15^{\circ}$.  The new pressure jump is estimated according to the following,
\begin{eqnarray}
\Delta {p^{new}} = \frac{{ - \tan \psi }}{{{{\left. {\frac{{\partial \tilde h}}{{\partial \tilde x}}} \right|}_{{\rm{midplane}}}}}}\sqrt {\gamma \Delta \rho g},
\end{eqnarray}We follow this procedure until we reach the convergence and the difference between two consecutive iterations becomes less than $1\%$. The pressure jump we found is $\Delta p = -0.2170 \frac{N}{m^2}$. The solution was also tested against the number of triangular meshes to confirm that the final result is independent of number of meshes in the domain of interest. Furthermore, to evaluate the principle curvatures, we evaluated higher order partial derivatives in each individual mesh according to,
 \begin{eqnarray}
 {h_{xx}} = \sum\limits_{i = 1}^3 {{h_i}\frac{{{\partial ^2}{\varphi _i}}}{{\partial {x^2}}}}
 \end{eqnarray}
 Thereafter, we calculate the principle curvature according to appropriate relations. The resulting deviatoric and mean curvature fields are shown in Fig.~\ref{Fig7}; the deviatoric curvature diverges near the corners, spanning values ranging from 400 $m^{-1}$ to 3000 $m^{-1}$ whereas mean curvature varies from 3 $m^{-1}$ to 5 $m^{-1}$. It is therefore safe to ignore capillary energies owing to mean curvature gradients.

Discussions with Dr. Lu Yao and Dr. Ravi Radhakrishnan are gratefully acknowledged. This work is partially supported by NSF grant CBET-$1133267$ and GAANN P200A120246 and MRSEC grant DMR11- $20901$.
\providecommand*{\mcitethebibliography}{\thebibliography}
\csname @ifundefined\endcsname{endmcitethebibliography}
{\let\endmcitethebibliography\endthebibliography}{}

%\bibliography{ref-2} %your .bib file
%\bibliographystyle{rsc} %the RSC's .bst file
\end{document}